\documentstyle[aps,twocolumn,epsfig]{revtex}
\begin{document}
\newcommand{\ve}[1]{\mbox{\boldmath $#1$}}
\twocolumn[\hsize\textwidth\columnwidth\hsize
\csname@twocolumnfalse%
\endcsname

\draft

\title {Vortices in Bose-Einstein condensates with anharmonic confinement}
\author{A. D. Jackson$^1$ and G. M. Kavoulakis$^2$}
\date{\today}
\address{$^1$Niels Bohr Institute, Blegdamsvej 17, DK-2100 Copenhagen \O,
             Denmark \\
         $^2$Mathematical Physics, Lund Institute of Technology, P.O.  Box
             118, S-22100 Lund, Sweden}
\maketitle

\begin{abstract}

We examine an effectively repulsive Bose-Einstein condensate of atoms,
that rotates in a quadratic-plus-quartic trapping potential. We investigate
the phase diagram of the system as a function of the angular frequency of 
rotation and of the coupling constant, demonstrating that there are phase 
transitions between multiply- and singly-quantized vortex states. The
derived phase diagram is shown to be universal and exact in the limits of 
small anharmonicity and weak coupling constant.

\end{abstract}

\vskip0.5pc]

The behavior of Bose-Einstein condensates of alkali-metal atoms
under rotation has become a very interesting topic in recent years,
since these systems are superfluid and the only way to rotate them
is by creating quantized vortex states.  The experimental observation
of vortex states in a two-component system was reported
by Matthews {\it et al.} \cite{JILA}, while Madison {\it et al.}
\cite{Madison} observed vortex states in a stirred one-component
Bose-Einstein condensate.  Many more experimental studies on rotating
condensates have been performed recently
\cite{VortexLatticeBEC,HaljanCornell,JILAgiant}.

If, in a harmonic trapping potential, the frequency of rotation
becomes equal to the trap frequency, the centrifugal force exactly
cancels the restoring force and the atoms fly apart. In other words,
the trap frequency sets an upper limit for the frequency of rotation
of the cloud. To overcome this difficulty and also to examine the
various phases which have been predicted theoretically
\cite{Lundh,Fetter,tku,FG,KB,Emil,SPB,AA}, Bretin {\it et al.}\,\cite{Dal}
recently examined experimentally a rotating Bose-Einstein condensate in an
anharmonic potential having the form
\begin{equation}
   V(\rho) = \frac 1 2 M \omega^2 \rho^2 [1 + \lambda (\frac {\rho} {a_0})^2].
\label{anh}
\end{equation}
Here, $\rho$ is the cylindrical polar coordinate, $M$ is the atomic mass,
$\omega$ is the frequency of the harmonic trapping potential,
$a_0 = (\hbar/M \omega)^{1/2}$ is the oscillator length, and $\lambda$
is a dimensionless constant which in the experiment of
Ref.\,\cite{Dal} was very small, on the order of $10^{-3}$.

In what follows we investigate the behavior of a gas that is
trapped in a potential of the form of $V$ and is under rotation.
In reality, the gas is also trapped in the $z$ direction.  However since
it rotates around this axis, we neglect the trapping in this direction and
assume that the gas has a constant density per unit length
$\sigma = N/Z$, where $N$ is the number of atoms and $Z$ is the width
of the condensate. In addition we assume that the degrees of freedom along
the $z$ axis are frozen so that our problem becomes two-dimensional.
In this case and for a purely harmonic potential, $\lambda = 0$,
the single-particle energy levels are given by
\begin{eqnarray}
  \epsilon = (2 n_r + |m| + 1) \hbar \omega,
\label{energy}
\end{eqnarray}
where $n_r$ is the radial quantum number, and $m$ is the quantum number
corresponding to the angular momentum.  Since $\epsilon$ grows linearly
with the angular momentum, any superposition of states with different
values of $m$ (and total angular momentum $m_0$) will have the same energy
as the pure state with $m = m_0$.  Interactions split this degeneracy.  As
shown in Refs.\,\cite{Rokhsar,KMP}, repulsive interactions always favor a
superposition of states.  The argument is very much like that in the
tight-binding model, where delocalization of the wavefunction lowers the
energy.  In practice, this implies that a singly-quantized vortex state
always has a lower energy than a multiply-quantized vortex in a
harmonic potential.

This picture changes drastically when $\lambda > 0$. The single-particle
energy levels grow faster than linearly with $m$ in this case. As a result,
for sufficiently weak interactions, a multiply-quantized vortex state
has lower energy than a singly-quantized vortex with the same circulation.
As the coupling constant increases, however, it eventually becomes
energetically favorable for the system to spread its angular
momentum to additional states, and multiply-quantized vortex states
break into singly-quantized vortices. We demonstrate this phase
transition explicitly below.

In the following, we assume that the interaction is described
by a contact potential of the form
\begin{eqnarray}
   V_{\rm int} =
   \frac 1 2 U_{0} \sum_{i \neq j} \delta({\bf r}_{i} - {\bf r}_{j}),
   \label{v}
\end{eqnarray}
where $U_0 = 4 \pi \hbar^2 a/M$ is the strength of the effective
two-body interaction with $a$ equal to the scattering length for
atom-atom collisions. We assume that the interaction is repulsive,
$a > 0$. Attractive interactions in anharmonic potentials have been
studied in Ref.\,\cite{Emil}.

We can estimate the interaction strength required at the transition
for a given value of the anharmonicity coefficient, $\lambda$, by noting
that the characteristic energy scale associated with the anharmonicity
must be comparable to the interaction energy. (This argument, valid when
the number of vortices is of order $1$, will be generalized below.)  The
first scale is of order $\lambda \hbar \omega$, and the second is of order
$\alpha n U_0$, where $n$ is the characteristic atom density, and $\alpha$
is a dimensionless constant of order $10^{-1}$.  (See, e.g., Fig.\,1 of
Ref.\,\cite{KMP}.)  If $\lambda \hbar \omega$ is to be $\approx \alpha n U_0$,
the effective coupling constant, $\sigma a$, must be $\approx \lambda/\alpha$.
For $\lambda \sim 10^{-3}$, we conclude that $\sigma a$ must be less than
$1$.

Therefore, provided that $\lambda$ and $\sigma a$ are smaller than unity,
we consider the eigenstates of the harmonic potential with zero radial
excitations and angular momentum $m \hbar$ ($m$ is positive for simplicity)
as our unperturbed states,
\begin{equation}
    \Phi_m({\rho, \phi}) = \frac 1 {(m! \pi a_0^2 Z)^{1/2}}
       \left( \frac {\rho}{a_0} \right)^{m} e^{i m \phi}
        e^{-\rho^{2}/2 a_0^2},
\label{phim}
\end{equation}
where $\phi$ is the angle in cylindrical polar coordinates.
The expectation value of  $KE + V$ in the state $\Phi_m$, where
$KE$ is the kinetic energy, $- \hbar^2 \nabla^2 / 2 M$, is
\begin{equation}
   \langle \Phi_m | KE + V | \Phi_m \rangle = \hbar \omega
   [m + \frac \lambda 2 (m+1) (m+2)]
\label{exp}
\end{equation}
measured with respect to the zero-point energy, $\hbar \omega$.  (All
energies will henceforth be measured with respect to $\hbar \omega$.)
Equation (\ref{exp}) shows explicitly the faster-than-linear (i.e., quadratic)
dependence of the energy on $m$.

Let us now incorporate the interactions. We use perturbation theory
to determine the interaction energy per particle in the state $\Phi_m$,
\begin{eqnarray}
\frac 1 N \langle \Phi_m | V_{\rm int} | \Phi_m \rangle &=&
 \frac {N (N-1)} {N} \frac {U_0} 2 \int |\Phi_m|^4 d^3 r
\nonumber \\
  &=& \hbar \omega \sigma a
  \frac {(2 m)!} {2^{2m} (m!)^2}.
\label{intexp}
\end{eqnarray}
Combining Eqs.\,(\ref{exp}) and (\ref{intexp}) we obtain the total
energy per particle in $\Phi_m$,
\begin{equation}
  \frac 1 N \frac {E_m} {\hbar \omega} = m + \frac \lambda 2 (m+1) (m+2) +
    \sigma a \frac {(2 m)!} {2^{2m} (m!)^2}.
\label{totexp}
\end{equation}
In a rotating frame of reference with some angular velocity $\Omega$,
the energy is
\begin{equation}
  \frac 1 N \frac {E_m'} {\hbar \omega} = m (1 - \frac {\Omega} {\omega})
  + \frac \lambda 2 (m+1) (m+2) +
      \sigma a \frac {(2 m)!} {2^{2m} (m!)^2}.
\label{totexprot}
\end{equation}
The above expression gives the energy per particle in the rotating frame
in terms of the parameters $\Omega/\omega$, $\lambda$, and $\sigma a$, in
a given state $\Phi_m$.

From Eq.\,(\ref{totexprot}) we see that $E_1'$ becomes lower than $E_0'$ for
a critical frequency of rotation which is
\begin{equation}
   \Omega_1/\omega = 1 + 2 \lambda - \sigma a/2.
\label{t}
\end{equation}
This is actually the critical angular frequency for inducing rotation
in the cloud. For $\lambda = 0$, this result is in agreement with
Refs.\,\cite{Rokhsar,KMP}. As $\Omega$ increases further, states with
larger $m$ become energetically favorable. The critical value of $\Omega_m$
for a multiply-quantized vortex state to form with $m \hbar$ units of
angular momentum is given by
\noindent
\begin{figure}
\begin{center}
\epsfig{file=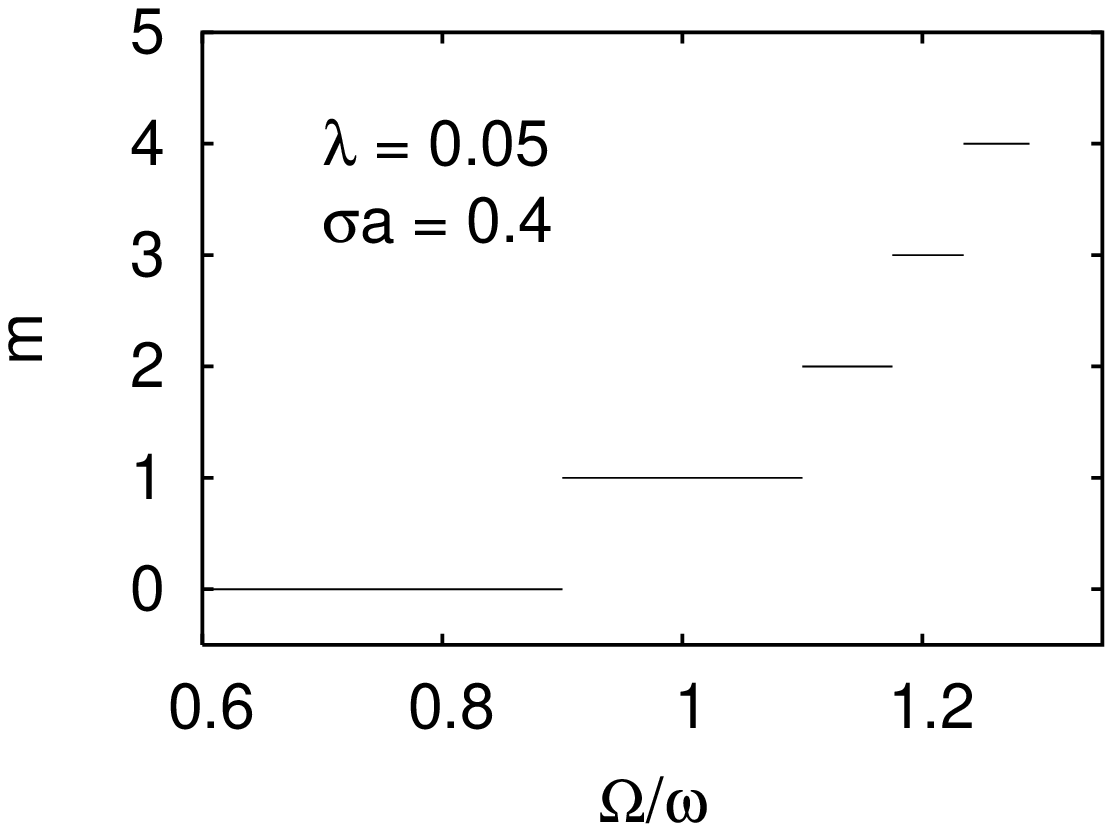,width=9.0cm,height=7.0cm,angle=0}
\vskip0.5pc
\begin{caption}
{The angular momentum, $m$, as function of the rotational frequency,
$\Omega/\omega$, as given by Eq.\,(\ref{gen}) with $\lambda = 0.05$ and
$\sigma a = 0.4$. As $\Omega/\omega$ increases, multiply-quantized
vortices form at the critical frequencies indicated in the graph.}
\end{caption}
\end{center}
\label{FIG1}
\end{figure}
\noindent
\begin{equation}
   \Omega_m / \omega = 1 + \lambda (m+1)
      - \sigma a \frac {(2 m - 2)!} {2^{2m-1} (m-1)! m!}.
\label{gen}
\end{equation}
A similar result was derived using a variational approach in
Ref.\,\cite{Emil} for an attractive Bose gas trapped in all three
directions. According to our model and for sufficiently weak interactions,
the angular momentum $m \hbar$ of the system is quantized and increases
stepwise at the frequencies given by Eq.\,(\ref{gen}), as shown in Fig.\,1
for $\lambda = 0.05$ and $\sigma a = 0.4$. For $m \gg 1$, $\Omega_m / \omega
\approx 1 + \lambda m$. Figure 2 shows the phase diagram of
multiply-quantized vortex states in the $\Omega/\omega$ -- $\sigma a$ plane,
as given by Eq.\,(\ref{gen}), for $\lambda = 0.05$.

We argued earlier that, as the strength of the interaction increases,
a multiply-quantized vortex state will split into singly-quantized vortices.
To attack this problem more generally, one can expand the order parameter
\cite{Rokhsar,KMP},
\begin{eqnarray}
   \Psi(\rho, \phi) = \sum_m c_{m} \Phi_{m}.
\label{exptrm}
\end{eqnarray}
Here the coefficients $c_{m}$ are (complex) variational parameters subject
to
\begin{eqnarray}
  \sum_{m} |c_{m}|^{2} = 1, \,\,\,\, \sum_{m} m |c_{m}|^{2} = l.
\label{cond1}
\end{eqnarray}
The first condition in simply the normalization, while the second implies
that the state $\Psi$ is assumed to have some angular momentum
per particle, $l \hbar$,
where $l$ is not necessarily quantized. After minimization of the energy
in the rotating frame with respect to the variational parameters $c_m$ and
with respect to $l$, one can determine $\Psi$ and $l = l (\Omega)$.

The phase transition we consider first appears in the case with $m=2$,
and this is the example we consider below. However our method can be
generalized to higher values of $m$. Let us therefore consider the
trial wavefunction
\cite{Rokhsar,KMP}
\begin{eqnarray}
   \Psi(\rho, \phi) = c_{0} \Phi_{0} + c_{2} \Phi_{2} + c_{4} \Phi_{4}.
\label{exptr}
\end{eqnarray}
The state $\Psi$ reduces to $\Phi_2$ when $|c_2|= 1, c_{0} = c_{4} = 0$
(i.e., it is a doubly-quantized vortex state), and it describes two
singly-quantized vortices when $c_{0} c_{4} \neq 0$ \cite{note}.
As we show below, there is a critical value of $\sigma a$ above which
$c_{0}$ and $c_{4}$ become nonzero. It is important to note that, as a
consequence of the spherical symmetry of the interaction, the phase boundary
between single and multiple quantization is given {\em exactly\/} by the
trial order parameter of Eq.\,(\ref{exptr}) so long as the states with
nonzero radial nodes, $n_r \neq 0$, can be neglected.  Equation (\ref{exptr})
can be extended to larger values of $m$ by considering the local stability
of the multiply-quantized state, $\Phi_m$, with respect to the admixture of
states with any angular momentum $m_1$ and $m_2 = 2m - m_1$.  For small $m$,
the leading instability is given by $m_1 = 0$.  These results will be
addressed in a future publication.

Calculating the energy per particle in the state $\Psi$ in the rotating
frame of reference we get,
\begin{eqnarray}
  \frac 1 N \frac {E'} {\hbar \omega} = l (1 - \Omega/\omega)
    + \lambda (|c_0|^2 + 6 |c_2|^2 + 15 |c_4|^2)
\nonumber \\
      + \sigma a (|c_0|^4 + |c_0|^2 |c_2|^2 + \frac 3 8 |c_2|^4 +
      \frac 1 4 |c_0|^2 |c_4|^2
\nonumber \\
      + \frac {35} {128} |c_4|^4
      + \frac {15} {16} |c_2|^2 |c_4|^2
      - \frac {\sqrt 6} 4 |c_0| |c_2|^2 |c_4|).
\label{totexptr}
\end{eqnarray}
Using Eqs.\,(\ref{cond1}), we can eliminate
$|c_0|$ and $|c_4|$ in Eq.\,(\ref{totexptr}). For a fixed $\lambda$, we
then vary $c_2$ and $l$ to minimize $E'$ for various values of
$\Omega/\omega$ and $\sigma a$ and thus investigate the phase diagram
in the $\Omega/\omega$ -- $\sigma a$ plane. The result for $\lambda = 0.05$
is shown as the dashed line in Fig.\,2 in the $m =2$ phase. Above this line 
there are two singly-quantized vortices with $|c_2| < 1$. Below this line 
there is a doubly-quantized vortex state, where $|c_2| = 1$. The actual
phase boundary is expected to lie lower than the dashed line because of
the variational nature of the present calculation (i.e., the neglect of
states with $n_r \ne 0$). Extending the same approach to higher values of
$m$, we derive the corresponding phase boundaries for $m = 3$, 4 and 5,
which are shown as dashed lines in Fig.\,2. The overall phase diagram
of Fig.\,2 resembles that obtained numerically by Lundh \cite{Lundh} for 
traps of some power law.

It is important to note that the transition seen here is continuous and
of second order.  Thus, it is sufficient to consider the local stability
of the states $\Phi_m$ in determining the phase boundary.  As $\sigma a$
increases, the separation between the two vortices increases continuously
in agreement with numerical simulations \cite{Lundh}.  Note also that the
form of the phase diagram is unaltered by changes in the strength of the
anharmonicity since $(1 - \Omega / \omega)$ and $\sigma a$ at 
\noindent
\begin{figure}
\begin{center}
\epsfig{file=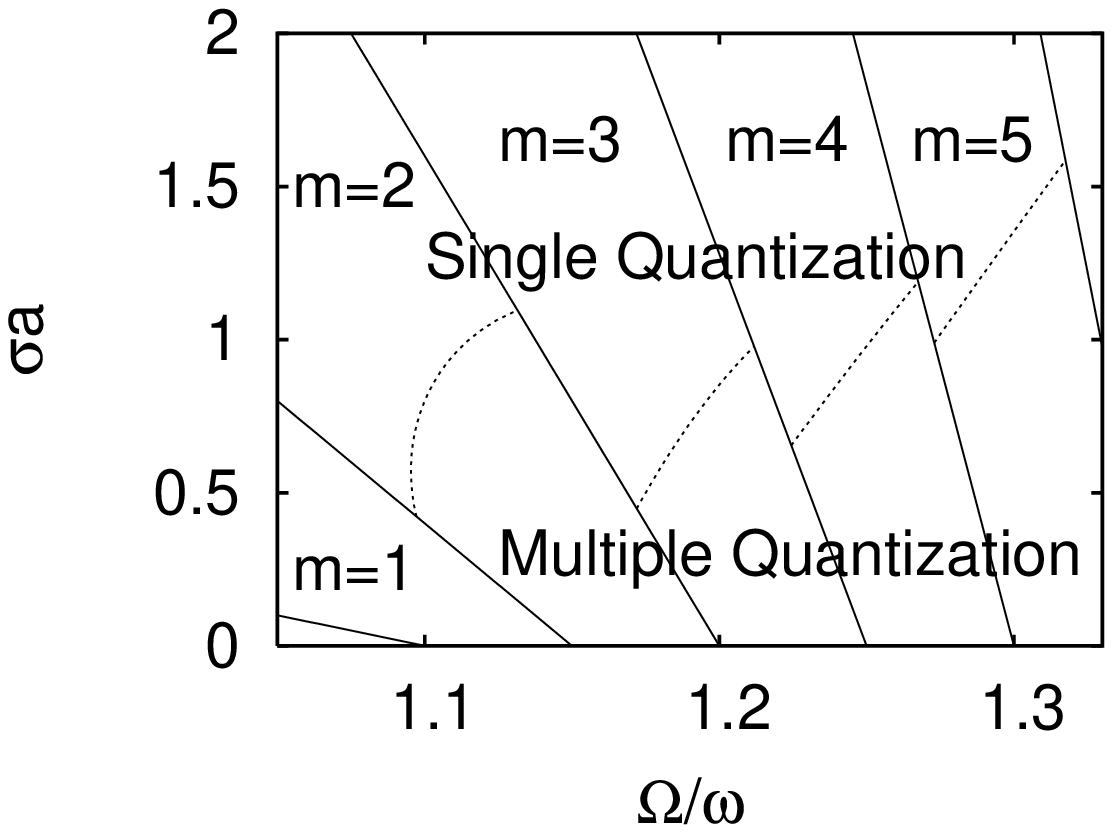,width=9.0cm,height=7.0cm,angle=0}
\vskip0.5pc
\begin{caption}
{The phase diagram of vortex states in a quadratic-plus-quartic potential
in the $\Omega/\omega$ -- $\sigma a$ plane for $\lambda = 0.05$. The solid
lines are the phase boundaries between multiply-quantized vortex states
with $m=0$, 1, 2, 3, 4, 5, and 6, as given by Eq.\,(\ref{gen}). The dashed lines
indicate the boundary between the phase of singly-quantized vortices,
and the phase of multiply-quantized vortex states.}
\end{caption}
\end{center}
\label{FIG2}
\end{figure}
\noindent
the phase boundary simply scale linearly with $\lambda$.  The phase diagram 
shown in Fig.\,2 is thus {\it universal} and {\it exact} in the limits of small 
anharmonicity and weak interactions.

Another important point is that in the phases of single
quantization, the function $l=l(\Omega)$ is not necessarily quantized.
Therefore, the graph of Fig.\,1 for higher values of $\sigma a$ will
be partly continuous and partly discontinuous. This is similar to
the case of harmonic confinememt \cite{Rokhsar} except for the fact
that, in the present problem $l$ does not diverge when
$\Omega \to \omega$ because of the presence of the quartic term
in the trapping potential.

To understand better the mechanism which gives rise to the phase
transition between single and multiple quantization, we return
to the trial order parameter of Eq.\,(\ref{exptr}) and consider the
extra constraint $l = 2$. Eliminating $|c_0|$ and $|c_4|$ in the energy,
we find in the laboratory frame that
\begin{eqnarray}
  \frac 1 N \frac {E} {\hbar \omega} &=&
     2 + 2 \lambda (4 - |c_2|^2)
     \nonumber \\
      + \frac {\sigma a} {512} [195 &-& (64 \sqrt 6 - 106) |c_2|^2
       + (64 \sqrt 6 - 109) |c_2|^4].
\label{totexptrf}
\end{eqnarray}
The right side of the above equation must be minimized with respect to
$|c_2|$. However, it is clear that the energy that results from the
anharmonicity is minimized for $|c_2|=1$, while the interaction energy
is minimized for $|c_2|^2 = (64 \sqrt 6 - 106)/ 2 (64 \sqrt 6 - 109)
\approx 0.53$, which is less than 1. It is essentially the balance between
these two terms that gives rise to the phase transition between multiple and
single quantization. Some trivial algebra shows that the
critical coupling constant above which the vortices separate is given by
\begin{eqnarray}
   (\sigma a)_c  = \frac {64 \lambda} {4 \sqrt 6 - 7} \approx 22.88 \lambda.
\label{cr}
\end{eqnarray}
The dashed curve in Fig.\,2 (for $m = 2$) saturates for larger $\Omega$
to the value of $\sigma a$ given by Eq.\,(\ref{cr}).  (In fact, it has
essentially reached this point at the highest value of $\sigma a$ shown.)
Since the solid line separating the $m = 2$ and $m = 3$ phases in Fig.\,2
cuts this dashed line at some $\sigma a \le (\sigma a)_c$, Eq.\,(\ref{cr})
provides an upper limit for $\sigma a$ for which there is a doubly-quantized
vortex state. In Ref.\,\cite{Lundh} the function $(\sigma a)_c =
(\sigma a)_c(\lambda)$ was calculated numerically (in Fig.\,4 of
\cite{Lundh}), and $(\sigma a)_c$ of Eq.\,(\ref{cr}) indeed lies above
this curve.  The discrepancy is less than 10 percent for the smallest
$\lambda$ shown, $\lambda = 0.01$, and grows as $\lambda$ increases.

We note that Ref.\,\cite{KB} has predicted the existence
of a triple point in the $\Omega/\omega$ -- $\sigma a$ plane, at which
a multiply-quantized vortex state, a vortex lattice, and a vortex lattice
with a hole coexist. The present study is far from this point.  However,
such a triple point would be indicated in our approach when the leading
instability is due to the admixture of states with $m_1 \ne 0$,
since $\Phi_{0}$ is the only component in the wavefunction that does not
vanish at $\rho = 0$.  Since the effective trapping potential,
$V - M \Omega^2 \rho^2/2$, has the familiar ``Mexican hat'' shape
for $\Omega/\omega > 1$, the contribution of $\Phi_{0}$ to $\Psi$ will
decrease with increasing $\Omega/\omega$.

Turning to the experiment of Ref.\,\cite{Dal}, for
$\lambda = 10^{-3}$ Eq.\,(\ref{cr}) implies a value of
$(\sigma a)_c \approx 0.02$, which is very small. Actually this experiment
is closer to the Thomas-Fermi limit of strong interactions \cite{BP}.
For a number of atoms $N = 3 \times 10^5$, $\omega = 2 \pi \times 65.6$ Hz,
$\omega_z = 2 \pi \times 11.0$ Hz, and $a = 53$ \AA, we find a transverse
radius of the cloud of $\approx 6.6$ $\mu$m, and a width in the $z$ direction
of $\approx 39.4$ $\mu$m. With these values $\sigma a$ turns out to
be $\approx 20$, which is much greater than $(\sigma a)_c$. This fact
explains why no multiply-quantized vortices were observed.  On the other
hand, tens of vortices were created in this experiment, and $(\sigma a)_c$
increases with the number of vortices \cite{KB}.
We can obtain a more general estimate for $(\sigma a)_c$ for the $m$-th
multiply-quantized vortex state as
\begin{eqnarray}
   (\sigma a)_c \sim \lambda \frac {2^{2m -1} m! (m+2)!} {2m!},
\label{crgen}
\end{eqnarray}
by equating the energy due to the quartic term, Eq.\,(\ref{exp}), with
the interaction energy, Eq.\,(\ref{intexp}). Equation (\ref{crgen}) is
indeed an increasing function of $m$. In Eq.\,(\ref{crgen})
$\lambda (m+1) (m+2) /2$ must be smaller than $m$, and $m$
should thus not exceed $2/\lambda$.  Self-consistency requires that
$(\sigma a)_c$ should not exceed unity. To consider a specific example with
$\lambda = 10^{-3}$ for a multiply-quantized vortex state with $m = 15$,
Eq.\,(\ref{crgen}) implies that $(\sigma a)_c \sim 0.94$, while
Eq.\,(\ref{gen}) gives a rotation frequency $\Omega_{15} \approx 1.01 \omega$.

In summary, we have examined a weakly-repulsive Bose-Einstein condensate
of atoms that is confined in a quadratic-plus-quartic trapping potential
and is under rotation. We have investigated the two phases that the system
exhibits -- multiply- and singly-quantized vortices -- as functions of
the frequency of rotation, the coupling constant and the anharmonicity.
We have also demonstrated that, by varying the coupling constant, the cloud
undergoes a phase transition from multiply- to singly-quantized vortex states.

\vskip0.5pc
\noindent Author GMK wishes to thank Gordon Baym, Emil Lundh, Ben Mottelson,
and Chris Pethick for useful discussions. He also acknowledges financial
support  from the Swedish Research Council (VR), and from the Swedish
Foundation for Strategic Research (SSF).


\begin{references}
\bibitem{note} There are actually four vortices in the state $\Psi$,
since $\Psi$ is $\sim ({\tilde z} - z_1) ({\tilde z} - z_1^*) 
({\tilde z} - z_2) ({\tilde z} - z_2^*) e^{- |{\tilde z}|^2/2}$, where
${\tilde z} = \rho e^{i \phi} = x + i y$, and therefore
it has four nodes on the $x$ - $y$ plane. However, two of them are
located far from the center of the cloud, where the density is
essentially zero \cite{KMP}.
\end{references}
\end{document}